\documentclass[prl,twocolumn,showpacs,preprintnumbers]{revtex4}
\usepackage{graphicx} 
\usepackage{dcolumn}
\usepackage{bm}
\usepackage{booktabs} 
\usepackage[colorlinks=true,linkcolor=red]{hyperref} 
\usepackage{memhfixc} 

\begin{document}

\title{Evidence of Structural Strain in Epitaxial Graphene Layers on 6H-SiC(0001)}
\author{Nicola Ferralis}
\affiliation{Department of Chemical Engineering, University of California,
 Berkeley, California 94720}
\author{Roya Maboudian}
\affiliation{Department of Chemical Engineering, University of California,
 Berkeley, California 94720}
\author{Carlo Carraro}
 \email{carraro@berkeley.edu}
\affiliation{Department of Chemical Engineering, University of California,
 Berkeley, California 94720}

\date{\today}

\begin{abstract}
The early stages of epitaxial graphene layer growth on the Si-terminated 6H-SiC(0001) are investigated by Auger electron spectroscopy (AES) and depolarized Raman spectroscopy. The selection of the depolarized component of the scattered light results in a significant increase in the C-C bond signal over the second order SiC Raman signal, which allows to resolve submonolayer growth, including individual, localized C=C dimers in a diamond-like carbon matrix for AES C/Si ratio of $\sim$3, and a strained graphene layer with delocalized electrons and Dirac single-band dispersion for AES C/Si ratio $>$6. The linear strain, measured at room temperature, is found to be compressive, which can be attributed to the large difference between the coefficients of thermal expansion of graphene and SiC. The magnitude of the compressive strain can be varied by adjusting the growth time at fixed annealing temperature. 

\end{abstract}


\maketitle

Graphene has been shown recently to possess many of the remarkable electronic properties of carbon nanotubes \cite{deHeer_2007}, while lending itself more readily to the planar paradigm of integrated-circuit fabrication processes. Epitaxial graphene (epigraphene) on silicon carbide (SiC) surfaces is emerging as an attractive process alternative to the painstaking layer-by-layer exfoliation of graphite crystals \cite{Berger_2004,  Berger_2006, Hass_2006, Hass_2007, deHeer_2007}. Epigraphene shares the key electron transport properties of free-standing exfoliated films \cite{Berger_2004,  Berger_2006, deHeer_2007}. However, in contrast to the exfoliated films, new features in the electronic structures appear in epitaxial graphene grown on the (0001) surface of 6H-SiC, whose origins remain controversial \cite{Bostwick_2007, Zhou_2007, Rotemberg_2008, Zhou_2008}. The novel electronic properties and the choice of the type of substrate are significant for the design of nonlinear devices and underscores the importance of substrate interactions in epitaxial films. This Letter is focussed on Si-terminated 6H-SiC(0001). 

The interaction of epigraphene with SiC substrate is mediated by a monolayer of C atoms, so called ``buffer layer", arranged in a honeycomb lattice, like graphene, but bonded in sp$^{3}$ configuration, with each atom forming a covalent bond to a Si atom beneath \cite{Hass_2007}. This buffer layer evolves from C-rich, high temperature surface reconstructions of 6H-SiC (0001) upon thermal desorption of Si atoms around T=1100 $^{\circ}$C. Annealing to higher temperature (1250 $^{\circ}$C) results in further desorption of Si, which promotes the formation of a second carbon layer, and deprives the original (topmost) carbon atoms of their covalent bonds to Si atoms, inducing sp$^{2}$ bonding configuration, i.e., into that of a graphene layer \cite{Forbeaux_1998, Berger_2004}. This paper presents a detailed investigation of the processes by which Si loss occurs, and its effect on the bonding, electronic and mechanical properties of the resulting carbon film. Our observations are made possible by the polarization analysis of the Raman signal from the sample surface. Although unpolarized Raman spectroscopy has been widely used in characterizing graphene layers \cite{Gupta_2006, Ferrari_2006, Graf_2006, Stampfer_2007, Pisana_2007, Casiraghi_2007, Calizo_2007, Ferrari_2007, Faugeras_2008}, this is, to the best our knowledge, the first time that {\it depolarized} Raman light has been employed to study epitaxial graphene films. This experimental approach is instrumental in resolving the epilayer signal over the background substrate contribution.

Specifically, we have been able to resolve [in terms of C/Si surface composition ratio determined by Auger electron spectroscopy (AES)] the formation of covalently bonded C=C dimers at the initial stages of carbonization, followed by the growth of submonolayer islands \cite{Hibino_2008, Ohta_2008} with delocalized electronic states, and finally by a fully developed graphene layer, which supports extended states with simple (single-band) Dirac states. Moreover, we report the observation of compressive strain in epitaxial graphene films at room temperature. This strain is a consequence of the strong pinning provided by the (0001) surface \cite{foot0} and results from the competition of two separate phenomena, which balance only partially. The first is the slight mismatch between the lattice constants of SiC and graphene at the synthesis temperature, which results in the epitaxial layer growing under slight tensile strain at high temperature. The second is the difference in thermal expansion coefficient between SiC (which contracts upon cooling) and graphene (which expands). We emphasize that graphene synthesis is a high temperature process. Thus, ambient temperature plays no special role in the phase diagram of the epitaxial film. From this consideration alone, it would be a remarkable coincidence indeed if room temperature were the point of zero strain for the epitaxial monolayer. As it happens, it is not. 

Graphene epilayers were produced by thermal annealing of the Si-face of 6H-SiC single crystals with (0001) orientation (CREE Research, Inc.). The specimens were cleaned ex-situ by repeated cycles comprising a UV-ozone treatment and a wet etch in concentrated hydrofluoric acid, followed by annealing under atmospheric pressure of 10\% H$_{2}$ in Ar at 1850 $^{\circ}$C, to remove surface scratches and to leave a regularly stepped surface \cite{Hass_2006}. This treatment produced a sharp $(3\times 3)$ low-energy electron diffraction (LEED) pattern of the surface. The samples were heated in ultrahigh vacuum by electron bombardment from the backside and cooled down to room temperature with cooling rate of $\le$ 1 Ks${}^{-1}$. At room temperature a threefold symmetric LEED pattern was observed, indicative of graphene formation, superimposed on the (6$\sqrt 3\times 6\sqrt 3) R30^{\circ}$ pattern of the SiC surface reconstruction \cite{Berger_2004, Forbeaux_1998}. Epigraphene film thickness was controlled by selecting a specific annealing temperature \cite{Berger_2004, deHeer_2007}. Following annealing, the surface C:Si ratio was obtained by AES \cite{Berger_2004,  deHeer_2007}. Raman spectra of the samples were acquired {\it ex situ} in a JYHoriba LabRAM spectrometer in backscattering configuration (excitation line: 632.817 nm, laser power at the sample: $\le$4 mW, beam spot: $\sim$1 $\mu$m). For each individual sample, Raman spectra were acquired in several different locations. No significant lineshifts due to layer non-uniformities were detected, within the experimental resolution.

While Si atom loss and average elemental surface composition are accurately monitored by AES, bonding configuration is best determined by vibrational spectroscopy. 
Raman scattering is a well-developed technique to investigate bonding and structure of carbonaceous materials \cite{Reich_2004, Ferrari_2001} and in some cases, even electronic properties, owing to the strongly resonant character of its cross section in graphite \cite{Thomsen_2000, Saito_2001}. However, the SiC substrate vibrations overwhelm the unpolarized signal from carbon-rich surface reconstructions. This happens because C=C bond vibrations are in the range of 1550-1650 cm$^{-1}$, where they overlap with the strong second order Raman signal of SiC \cite{Nakashima_1997}. Here, however, we use the depolarized scattering configuration, which is justified by considerations of crystal symmetry \cite{foot1}.

\begin{figure}
\includegraphics{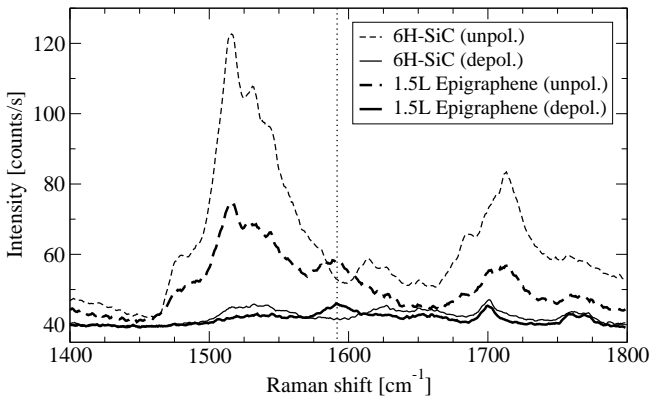}

\caption{Effects of the scattering depolarization on the second order scattering from the 6H-SiC(0001) bulk, and with 1.5L epigraphene. The vertical dotted line indicates the position of the epigraphene G peak, $\sim$1592 cm$^{-1}$.}

\end{figure}

The unpolarized Raman spectra of clean 6H-SiC(0001) and with 1.5 layers of epitaxial graphene (AES C/Si ratio=6.4) are shown in Fig.~1 (dashed lines). In this Letter, we refer to the 0th layer as the buffer. The SiC spectrum displays a strong second order signal in the region of interest (1500-1700 cm$^{-1}$) in which the first order graphene peaks show up. The corresponding depolarized spectra of the clean 6H-SiC(0001) and 1.5 layers of epitaxial graphene are shown in Fig. 1 (solid lines) confirming the expectations of a large reduction in second order signal. On the other hand, the zone center optical phonon line of graphene is invariant under rotation of the incident beam polarization with respect to the polarization of the analyzer, so that the depolarized signal is about one half as intense as the unpolarized one. Thus, the detection of submonolayer sp$^{2}$ bonding by Raman scattering on 6H-SiC (0001) becomes a distinct possibility in depolarized scattering configuration. 

\begin{figure}
\includegraphics{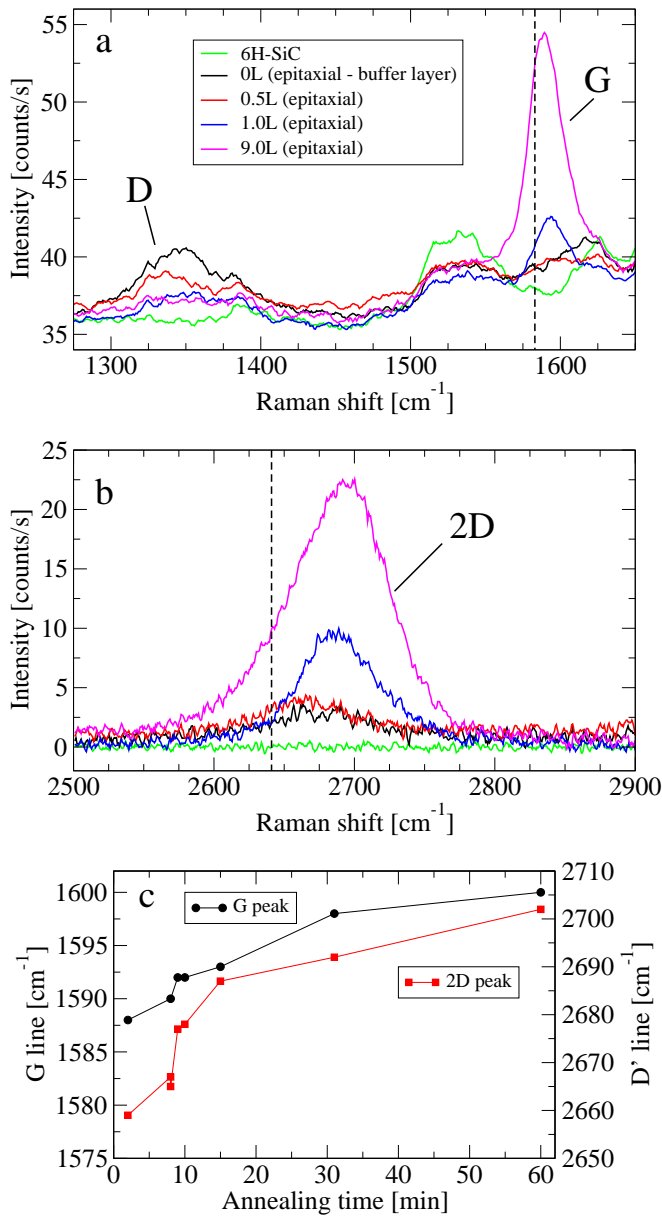}

\caption{Depolarized Raman spectra of the a) D and G peak and b) 2D peak regions. The average thicknesses of the epigraphene films (grown for the same annealing time, 8 min) are determined by the C:Si AES intensity ratio. Dotted lines in panels a and b correspond to the position of the G and 2D peak respectively, for 1 layer exfoliated graphene \cite{Ferrari_2006}. The evolution of G and 2D Raman shifts lines corresponding to $\sim$1 layer are plotted as function of the annealing times (panel c).} 

\end{figure}

Figures 2a, and 2b show the depolarized Raman spectra of a clean SiC sample, and of a series of epigraphene films on SiC samples obtained by annealing at various temperatures. The annealed samples are identified by their respective  C/Si ratios measured by AES, which are 2.7$\pm$0.1, 4.2$\pm$0.1, 7.0$\pm$0.2, and 11.0$\pm$0.2. These ratios correspond to 0 (buffer), 0.5, and 1 and 1.5 carbon layers \cite{deHeer_2007}. In Fig. 2a, the Raman spectra are compared in the region 1250-1750 cm$^{-1}$. The depolarized SiC spectrum is nearly featureless, except for a broad feature around 1500-1550 cm$^{-1}$, which make up the background signal common to all epitaxially grown samples. For reference, we marked with a dashed line the position of $\Gamma$ point optical phonon (G line) at 1582 cm$^{-1}$, for a monolayer of exfoliated graphene \cite{Ferrari_2006}. In addition to the G line, a disorder-induced peak (D line, around  1330 cm$^{-1}$) is visible in all graphene samples. This resonant D peak is particularly sensitive to the excitation frequency (the data shown here were acquired at 632.817 nm incident wave length). It appears much less pronounced (and blueshifted) at green or blue excitation source frequencies. 

The spectra of annealed SiC samples show several remarkable features. The G line (Fig. 2a) is found only in those samples for which the C/Si AES ratio is higher than 6, corresponding approximately to one carbon layer (in addition to the buffer layer) and thicker, whereas the buffer layer sample (0L) displays instead a band around 1620 cm$^{-1}$. This is close to the vibrational frequency of isolated C=C dimers in a diamond-like (sp$^{3}$-bonded) layer \cite{Ferrari_2001}. Mirroring the onset of the G line and its increasing intensity with increasing C/Si AES ratio, the amplitude of the D peak decreases very rapidly. 

The presence of the vibrational band around 1620 cm$^{-1}$ signals the incipient loss of subsurface Si atoms. An excess of dangling bonds develops in the topmost carbon layer, which begins to reconstruct locally and bond in sp$^2$ configuration. The shift of the intensity to lower wave numbers (in the G line range, 1580-1600 cm$^{-1}$) for the 0L sample signals the formation of delocalized orbitals, i.e., of finite sized island of graphene. This scenario is confirmed by the appearance, at about the same coverage, of a strong and relatively narrow 2D peak (see Fig. 2b), indicative of an electronic structure dominated by a single band with Dirac dispersion \cite{Faugeras_2008}. With continuing Si desorption, and increasing carbon coverage, the G and 2D peaks become more intense, whereas the D peak intensity drops. Thus, the increasing thickness of the graphene film is accompanied by healing of the defects in the topmost layers. 

A remarkable feature in the spectra is a blue shift, with respect to exfoliated graphene, which is observed in both the G {\it and} the 2D lines in a ratio of about 1:3, for a specific layer thickness. Equally remarkable is the observation that graphene growth time (i.e., the time for which the sample is held at the synthesis temperature) affects the magnitude of the blue shifts of both lines, as shown in Fig.~2c for $\sim$1 layer. The simultaneous shifts in both G and 2D line shifts can only be attributed to a uniform (hydrostatic) compressive strain. The mode-dependent relation between peak shift $\Delta\omega$ and strain tensor $\epsilon$ is given by 
\begin{equation}
{\Delta\omega\over\omega}=-\gamma_m{\rm tr}\epsilon_{ij},
\label{eq1}
\end{equation}
where $\gamma_m$ is the mode Gr\"uneisen parameter. For graphene, $\gamma_G\approx1.8$ and $\gamma_{D}\approx2.7$ \cite{Mounet_2005}, implying a ratio in blue shifts of the G and 2D peaks of about 1:2.5, in good agreement with our measurements \cite{foot2}.

We exclude any significant contribution to the observed shifts induced by charge. Epitaxial graphene on 6H-SiC is known to be negatively charged, with excess electron density of about $1.3\times10^{13}$ cm${}^{-2}$. In exfoliated graphene, this doping level induces a blue shift of the G line of about 7 cm$^{-1}$, but does {\it not} appreciably shift the 2D peak \cite{Das_2008}. The simultaneous shift of both lines, as well as the magnitude of the shifts observed at long annealing times, rule out charge transfer as the primary cause \cite{foot3}.

The origin of uniform compressive strain in epitaxial graphene is attributed to the large difference in the coefficients of linear  thermal expansion between SiC ($\alpha_{\rm SiC}$, measured \cite{Slack_1975}) and graphene ($\alpha_{\rm gr}$, calculated \cite{Mounet_2005}). This difference $\Delta\alpha(T)$ is nearly constant between room temperature (RT) and the graphene synthesis temperature, $T_s\approx 1250^{\circ}$C. If the epitaxial film grew in mechanical equilibrium with the SiC surface, that is, as a stress-free monolayer commensurate with the 6$\times \sqrt 3$-reconstructed SiC surface at $T_s$, a large compressive strain would develop in the film upon cooling, since SiC contracts on cooling, while graphene expands:
\begin{equation}
{\frac{1}{1-\epsilon}}=\exp\left[\int_{RT}^{T_{s}}dT' \Delta\alpha(T')\right].
\label{eq2}
\end{equation}
The calculated linear compressive strain at RT is about 0.8\%. By equation~(\ref{eq1}), this strain corresponds to a shift of about 22 cm$^{-1}$ for the G line, which is close to the values we observe for long annealing times. We conclude that long annealing times, in excess of an hour, are needed to produce graphene layers that are in mechanical equilibrium with the substrate at the growth temperature. Shorter times result in films that are formed under tensile stress \cite{foot4}.

In summary, we have investigated the epitaxial growth of graphene monolayers on the pinning 6H-SiC(0001) surface by AES and depolarized Raman spectroscopy. We have observed the thin graphene films at room temperature under residual compressive strain. The strain results from a large difference in the thermal expansion coefficients of graphene and SiC, and can be tuned by varying the growth time, between the theoretical maximum of about 0.8\% and an empirical minimum value of 0.1\%. 

\begin{acknowledgments}
We acknowledge useful discussions with V. Srinivasan and J.C. Grossman. This work was supported by the National Science Foundation Grants EEC-0425914 and CMMI-0825531, and by DARPA N/MEMS Science and Technology Fundamentals Center on Interfacial Engineering for MEMS.
\end{acknowledgments}


\begin{thebibliography}{40} 

\bibitem{deHeer_2007} W.A. de Heer, C. Berger, X. Wu, P.N. First, E.H. Conrad, X. Li, T. Li, M. Sprinkle, J. Hass, M.L. Sadowski, M. Potemski and G. Martinez, Solid State Comm. {\bf 143}, 92 (2007).

\bibitem{Berger_2004} C. Berger, Z. Song, T. Li, X. Li, A.Y. Ogbazghi, R. Feng, Z. Dai, A.N. Marchenkov, E.H. Conrad, P.N. First and W.A. de Heer, J. Phys. Chem. B {\bf 108}, 19912 (2004).

\bibitem{Berger_2006} C. Berger, Z. Song, X. Li, X. Wu, N. Brown, C. Naud, D. Mayou, T. Li, J. Hass, A.N. Marchenkov, E.H. Conrad, P.N. First and W.A. de Heer, Science {\bf 312}, 1191 (2006).

\bibitem{Hass_2006} J. Hass, R. Feng, T. Li, X. Li, Z. Zong, W.A. de Heer, P.N. First, E.H. Conrad, C.A. Jeffrey and C. Berger, Appl. Phys. Lett. {\bf 89}, 143106 (2006).

\bibitem{Hass_2007} J. Hass, R. Feng, J.E. Mill{\' a}n-Otoya, X. Li, M. Sprinkle, P. N. First, W.A. de Heer, E.H. Conrad and C. Berger, Phys. Rev. B {\bf 75}, 214109 (2007).

\bibitem{Bostwick_2007} A. Bostwick, T. Ohta, Th. Seyller, K. Horn and E. Rotenberg, Nat. Phys. {\bf 3}, 36 (2007).

\bibitem{Rotemberg_2008} E. Rotemberg, A. Bostwick, T. Ohta, J.L McChesney, Th. Seyller, K. Horn, Nat. Mat. {\bf 7}, 258 (2008).           

\bibitem{Zhou_2007} S. Y. Zhou, G.-H. Gweon, A.V. Fedorov, P.N. First, W.A. de Heer, D.-H. Lee, F. Guinea, A.H. Castro Neto and A. Lanzara, Nat. Mat. {\bf 6}, 770 (2007).

\bibitem{Zhou_2008} S.Y. Zhou, D.A. Siegel, A.V. Fedorov, F. El Gabaly, A.K. Schmid, A.H. Castro Neto, D.-H. Lee, A. Lanzara, Nat. Mat. {\bf 7}, 259 (2008).       

\bibitem{Forbeaux_1998} I. Forbeaux, J.-M. Themlin and J.-M. Debever, Phys. Rev. B {\bf 58}, 16396  (1998) and references therein.

\bibitem{Gupta_2006} A. Gupta, G. Chen, P. Joshi, S. Tadigadapa and P.C. Eklund, Nano Lett. {\bf 6}, 2667 (2006). 

\bibitem{Ferrari_2006} A.C. Ferrari, J.C. Meyer, V. Scardaci, C. Casiraghi, M. Lazzeri, F. Mauri, S. Piscanec, D. Jiang, K.S. Novoselov, S. Roth and A.K. Geim, Phys. Rev. Lett. {\bf 97}, 187401 (2006).

\bibitem{Graf_2006} D. Graf, F. Molitor, K. Ensslin, C. Stampfer, A. Jungen, C. Hierold and F. Wirtz, Nano Lett. {\bf 6}, 238 (2006).

\bibitem{Stampfer_2007} C. Stampfer, F. Molitor, D. Graf, K. Ensslin, A. Jungen, C. Hierold and F. Wirtz, Appl. Phys. Lett. {\bf 91}, 241907 (2007).

\bibitem{Pisana_2007} S. Pisana, M. Lazzeri, C. Casiraghi, K.S. Novoselov, A.K. Geim, A.C. Ferrari and F. Mauri, Nat. Mat. {\bf 6}, 198 (2007).

\bibitem{Casiraghi_2007} C. Casiraghi, S. Pisana, K.S. Novoselov, A.K. Geim, A.C. Ferrari, Appl. Phys. Lett. {\bf 91}, 233108 (2007).

\bibitem{Calizo_2007} I. Calizo, A.A. Balandin, W. Bao, F. Miao and C.N. Lau, Nano Lett. {\bf 7}, 2645  (2007).

\bibitem{Ferrari_2007} A.C. Ferrari, Sol. State Comm. {\bf 143}, 47 (2007).

\bibitem{Faugeras_2008} C. Faugeras, A. Nerri\`ere, M. Potemski, A. Mahmood, E. Dujardin, C. Berger and W. A. de Heer, Appl. Phys. Lett. {\bf 92}, 011914 (2008).

\bibitem{Hibino_2008} H. Hibino, H. Kageshima, F. Maeda, M. Nagase, Y. Kobayashi and H. Yamaguchi, Phys. Rev. B {\bf 77}, 075413 (2008).

\bibitem{Ohta_2008} T. Ohta, F. El Gabaly, A. Bostwick, J.L. McChesney, K.V. Emtsev, A.K. Schmid, Th. Seyller, K. Horn, E. Rotemberg, New Jour. Phys. {\bf 10}, 023034 (2008).

\bibitem{foot0} In contrast, graphene epitaxial layers are not pinned to the C-terminated 6H-SiC (000$\bar 1$) face, on which the graphene LEED pattern consists of rings rather than sharp spots \cite{Hass_2006}. 

\bibitem{Reich_2004} S. Reich and C. Thomsen, Phil. Trans. R. Soc. Lond. A {\bf 362}, 2271 (2004).

\bibitem{Ferrari_2001} A.C. Ferrari and J. Robertson, Phys. Rev. B {\bf 64}, 075414 (2001).

\bibitem{Thomsen_2000} C. Thomsen, S. Reich, Phys. Rev. Lett. {\bf 85}, 5214 (2000).

\bibitem{Saito_2001} R. Saito, A. Jorio, A. G. Souza Filho, G. Dresselhaus, M.S. Dresselhaus and M.A. Pimenta, Phys. Rev. Lett. {\bf 88}, 027401 (2001).  

\bibitem{Nakashima_1997} S. Nakashima, H. Harima, Phys. Stat. Sol. A {\bf 162}, 39 (1997).

\bibitem{foot1} The second order signal of SiC is due in large part to overtones and/or combinations involving zone center LO phonons [J.C. Burton, L. Sun, F.H. Long, Z.C. Feng, I.T. Ferguson, Phys. Rev. B {\bf 59}, 7282  (1999)]. The selection rules for the symmetry group $C_{6v}$ forbid this mode in depolarized scattering configuration.

\bibitem{Mounet_2005} N. Mounet, N. Marzari, Phys. Rev. B {\bf 71}, 205214 (2005).

\bibitem{foot2} The same behavior is observed for thicker layers, suggesting some degree of strain, probably due to pinning at domain edges.

\bibitem{Das_2008} A. Das, S. Pisana, B. Chakraborty, S. Piscanec, S.H. Saha, U.V. Waghmare, K.S. Novoselov, H.R. Krishnamurthy, A.K. Geim, A.C. Ferrari, and A.K. Sood, Nat. Nanotechnology {\bf 3}, 210 (2008).

\bibitem{foot3} Note that the other striking feature of excess charge in Raman spectra of exfoliated graphene, i.e., a narrowing of the G line, is in contradiction with our data, further ruling out excess charge as the cause of the spectral shifts observed here \cite{Casiraghi_2007}.

\bibitem{Slack_1975} G.A. Slack, S.F. Bartram, J. Appl. Phys. {\bf 46}, 89 (1975).

\bibitem{foot4} Unlike the annealing temperature (which controls film thickness) and annealing time (which controls film stress), we have found no film property dependence upon sample cooling rate. 

\end{thebibliography}
\end{document}